# Compressed Sensing in Multi-Hop Large-Scale Wireless Sensor Networks Based on Routing Topology Tomography

Yimei Li and Yao Liang

**Abstract**—Data acquisition from a multi-hop large-scale outdoor wireless sensor network (WSN) deployment for environmental monitoring is full of challenges. This is because the severe resource constraints on small battery-operated motes (e.g., bandwidth, memory, power, and computing capacity), the big data acquisition volume from the large-scale WSN, and the highly dynamic wireless link conditions in an outdoor communication environment. We present a novel compressed sensing approach which can recover the sensing data at the sink with high fidelity when very few data packets are collected, leading to a significant reduction of the network transmissions and thus an extension of the WSN lifetime. Interplaying with the dynamic WSN routing topology, the proposed approach is efficient and simple to implement on the resource-constrained motes without a mote's storing of any part of the random projection matrix, as opposed to other existing compressed sensing based schemes. We propose a systematic method via machine learning to find a suitable representation basis, for any given WSN deployment and data field, which is both sparse and incoherent with the random projection matrix in the compressed sensing for data acquisition. We validate our approach and evaluate its performance using a real-world multi-hop WSN testbed deployment in situ. The results demonstrate that our approach significantly outperforms existing compressed sensing approaches by reducing data recovery errors by an order of magnitude for the entire WSN observation field, while drastically reducing wireless communication costs at the same time.

## 1 Introduction

Wireless Sensor Networks (WSNs), comprised of spatially distributed sensor nodes, are being increasingly deployed for continuous monitoring and sensing of physical variables of our world [1-4]. One of the critical challenges in large-scale outdoor WSN deployments is energy consumption, since outdoor sensor nodes are mainly operated by battery power. Motivated by the breakthrough of compressed sensing (CS) [5, 6], CS based approaches for WSN data collection have gained increasing attention from the research communities (e.g., [7-14]). However, existing CS methods for WSNs are facing the following major difficulties in practice: First, how to effectively and efficiently interplay with WSN routing so that per-packet routing path can be exploited as a random projection in CS measurement matrix to further reduce nodes' transmissions? Second, how to design a suitable representation basis for real-world signals that has good sparsification and incoherence with the measurement matrix for applying CS to large-scale WSN data acquisition? As Quer et al. put it, "finding a suitable transformation with good sparsification and incoherence properties remains an open problem" [7]. Furthermore, existing CS approaches for multi-hop WSN data acquisition are only evaluated by numerical simulations with the assumptions of some routing models. While useful, numerical simulations alone are not adequate. The lack of validation in multi-hop WSN deployments in situ operated in real-world dynamic communication environments hinders any deep understanding of CS approaches for wireless big data and their meaningful comparison. Therefore, the need of practical validation and evaluation of CS approaches in real WSNs in situ is also urgent.

The objective of this work is to address the above challenges in the emerging Internet of Things and attempt to fill the gap. We present a practical and efficient CS solution for large-scale real-world WSN data acquisition, and focus on the joint compression and routing in outdoor multi-hop WSNs in situ where the communication environment is highly dynamic and harsh. The major contributions of this paper are as follows:

• We present a novel compressed sensing approach for multi-hop large-scale dynamic WSNs in situ for data acquisition based on network routing topology tomography.

• We propose a systematic method, based on graph wavelets via deep learning, to find an optimized representation basis which is extremely sparse and also

• Y. Li is with the Department of Computer and Information Science, Indiana University Purdue University Indianapolis (IUPUI), IN 46202. E-mail: liyim@imail.iu.edu.
• Y. Liang is with the Department of Computer and Information Science, IUPUI, IN 46202. E-mail: yaoliang@iupui.edu.

incoherent with the measurement matrix in our CS approach.

- We validate and evaluate our approach in an environmental multi-hop WSN deployment in a watershed, operating with TinyOS and an extended Collection Tree Protocol (CTP), in comparison with existing CS approaches. To the best of our knowledge, this work represents the first study on the CS approach for data acquisition conducted on a real-world outdoor WSN in situ with the deployed routing protocol and routing topology tomography.

The remainder of the paper is organized as follows. Section 2 presents related work. Section 3 overviews the mathematical background of CS theory. Section 4 presents our approach. In Section 5, we provide the validation and evaluation of our approach using a real outdoor WSN deployment in situ. Finally, Section 6 concludes the paper.

## 2 RELATED WORK

In the recent years, many research efforts have been pursued to incorporate CS into data collection schemes in WSNs (e.g., [7-14, 31, 32]). Traditional CS based approaches such as [8-10] do not exploit the knowledge about WSN routing topology but rely on the use of dense measurement matrices, resulting in high transmission costs and storing a part of measurement matrix in each resource-constrained sensor node [e.g., 8, 10]. Wang et al. [32] studied CS based on sparse random projections for WSN data querying without interaction with routing. While the approach of [32] could reduce WSN transmission costs for data nodes compared to the CS approaches based on dense measurement matrices, it does not solve the problem of storing a part of the measurement matrix at each sensor node, and its performance would also be largely diminished in multi-hop WSNs. On the other hand, Quer et al. [7] studied the interplay of routing with compressive sensing in multi-hop WSNs, where the measurement matrix is defined according to the routing paths. However, the authors of [7] found the results of their work were unsatisfactory due to the difficulty to find a suitable representation basis for real signals, stating that "finding a suitable transformation with good sparsification and incoherence properties remains an open problem" for WSN data acquisition. The authors of [13] presented some theoretical analysis regarding the nonuniform random projection of CS. However, it is not clear if their analysis is applicable to the situation where the nonuniform random projection of CS projection is formed from practical WSN routing. Besides, in the approach of [13], each per-packet routing path is recorded in the data packet routed towards the sink, which is neither scalable nor efficient. For example, if a node identifier is two bytes (as in tinyOS), then for a WSN of the maximum path of J hops to the sink it would have to allocate 2(J-1) bytes in a data packet for its path recording overhead. This heavy overhead of path recording also increases energy and bandwidth consumptions for transmissions, reducing or eliminating the performance of data compression. Zheng et al. [14] propose a random walk algorithm for data gathering in multi-hop WSNs, the measurements are collected along the random walks before they are sent to the sink using shortest path routing. Therefore, the method proposed in [14] does not interplay with WSN routing. Due to the fact that it requires the length of each walk t=O(n/k) for each packet before routing to the sink, the method of [14] increases the WSN energy consumption due to the additional random walk transmissions. Another approach to compute projections is based on analog communications [31], where CS projections are simultaneously calculated by the superposition of radio waves and communicated directly from the sensor nodes to the sink via the air interface. This approach, however, requires analog communications for WSNs, which is in contrast to today's digital communications commonly used in WSN physical layer, such as IEEE 802.15.4 communication protocol. Firooz and Roy [23] studied network link delay estimation using CS via expander graphs when the routing matrix is predetermined; they demonstrated the fesibility of accurate estimation with bounded errors. Some other researches [11, 12] focused on temporal correlations in a sequence of samples taken by each sensor node in WSN. Besides, no published work so far validated CS performance in multi-hop WSNs through real experiments on WSN deployments in situ with actual routing protocol in operation.

## 3 COMPRESSED SENSING BACKGRAOUND

Compressed sensing is a breakthrough technique in signal processing [5, 6], CS theory asserts that for sparse or compressible signals, one can actually recover the original signals by using far fewer measurements or samples than required by the Nyquist rate. Consider an N-dimensional discrete sparse signal vector $x \in \mathbb{R}^N$, which is referred to as $k$-sparse if $x$ has no more than $k$ ($k \ll N$) nonzero items. Mathematically, the theory of CS has shown that if $x$ is sparse, under certain conditions, then it is possible to reconstruct the signal vector $x$ from $M$ measurements $y = [y^1, y^2, ..., y^M]^T$ with a quasi-random $M \times N$ *measurement matrix* $\Phi$, i.e., $y = \Phi x$, where $M$ ($k < M$) is much smaller than $N$. This can be achieved (with probability close to one) by solving the following optimization:

$$\min_x ||x||_p \quad \text{s. t.} \quad y = \Phi x, \qquad (1)$$

where $||x||_p$ ($p = 0, 1$) denotes $l_p$-norm of $x$. Often, a signal $x$ is not sparse but can be sparsely represented in an alternative domain. Specifically, if $x$ can be further written as $x = \Psi s$, for some $N \times N$ matrix $\Psi$, where $s$ is the $N \times 1$ coefficient vector in the $\Psi$-domain with $\|s\|_o = k$, the matrix $\Psi$ will be referred to as the *representation basis*. We have $y = \Phi\Psi s = \widetilde{\Phi} s$, where $\widetilde{\Phi} = \Phi\Psi$ is also quasi-random. Then the associated signal recovery problem is to determine $s$ for given measurements $y$ and the defined matrices $\Phi$ and $\Psi$:

$$\min_s \|s\|_p \quad \text{s. t.} \quad y = \widetilde{\Phi} s. \quad (2)$$

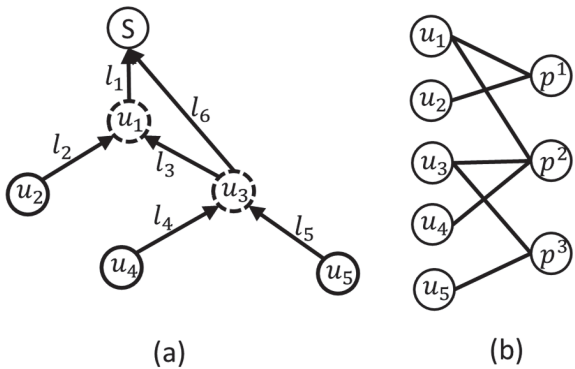

Fig. 1. (a) An illustration of sensor network upward routing topology for data collection: solid circles (the sink and leaves) are the boundary nodes and dash circles are the intermediate nodes. (b) Bipartite graph corresponding to given routing matrix in Eq. (4).

As $M$ is much smaller than $N$, this is an under-determined linear system. The reconstruction of the original signal $x$ is given by

$$x = \Psi s. \quad (3)$$

## 4 APPROACH

### 4.1 Problem Formulation

To minimize the number of transmissions, our compressed sensing approach for multi-hop WSN data collection, referred to as CSR (Compressed Sensing based on dynamic Routing topology tomography), closely interplays with the dynamic routing topology in a given WSN deployment. As a data packet is routed from its source node towards the sink, the sensor reading of each traversed node adds up along the path. Let a dynamic WSN for data collection be modeled as a directed acyclic graph $G(V, E)$, where $V$ is a set of $n$ nodes (the sink $S$ and $n-1$ sensor nodes), and $E$ is a set of edges. A directed edge $e(u, v)$, an ordered pair $(u, v) \in V \times V$, represents the wireless communication link from node $u$ to node $v$. Let $p^i = [u_0^i, u_1^i, ..., u_j^i, ... S]$ denote a routing path of packet $i$ from a source node $u_0^i$ to the sink $S$, which is a sequence of all nodes traversed along the route. For example, as shown in Figure 1, there are three data collection paths initiated from leaf nodes in a collection cycle:

$p^1 = [u_2, u_1, S],$

$p^2 = [u_4, u_3, u_1, S],$

$p^3 = [u_5, u_3, S].$

Let $\Phi$ denote the routing matrix corresponding to the set of paths $P_\Phi = \{p^1, p^2, p^3\}$. Then, for the sensor network shown in Figure 1(a), the routing matrix $\Phi$ for the given data collection cycle is as follows:

$$\Phi = \begin{matrix} p^1: u_2 \to S \\ p^2: u_4 \to S \\ p^3: u_5 \to S \end{matrix} \begin{bmatrix} u_1 & u_2 & u_3 & u_4 & u_5 \\ 1 & 1 & 0 & 0 & 0 \\ 1 & 0 & 1 & 1 & 0 \\ 0 & 0 & 1 & 0 & 1 \end{bmatrix} \quad (4)$$

A bipartite graph $B(V, P_\Phi, H)$ can be formed from a $G(V, E)$ with a bi-adjacency matrix $\Phi$, where V is the set of nodes in $G(V, E)$, and $H \subset V \times P_\Phi$ is a set of coupled elements from V to $P_\Phi$. Figure 1(b) represents the bipartite graph for the WSN in Figure 1(a) with the routing matrix $\Phi$.

Let $y_j^i$, carried by packet $i$, denote the aggregated compressed sensor reading measurement at node $u_j^i$ along the route $p^i$ towards the sink. We define the following in-network compressing operation for each data packet $i$:

$$y_0^i = \text{reading}(u_0^i), \quad j = 0, \quad (5)$$

$$y_j^i = y_{j-1}^i + \text{reading}(u_j^i), \quad j > 0, \quad (6)$$

where $y_j^i$ is computed on the fly at each intermediate node $j$ along the dynamic route towards the sink. In our approach, $M$ ($M \ll n-1$) data packets initiated from $M$ randomly selected source nodes of the WSN are collected in each data collection cycle, which carry $M$ compressed sensing measurements specified by (5) and (6), along their respective routing paths, denoted by $y = [y^1, y^2, ..., y^M]^T$. As one can see, each data collection routing path represents a random projection of the WSN data field in compressed sensing. In general, a routing path in an outdoor WSN is inherently random due to the highly dynamic wireless link conditions of the WSN. In addition, some WSN routing protocols could further induce more randomness in routing paths.

A critical issue of such a compressed sensing formed via WSN routing is how to obtain such dynamic routing path information at the sink. Since we consider realistic WSN deployments *in situ* under *time-varying* communication environments, where wireless links available a moment ago for a previous packet transmission may not be available for the current packet in a random way, such on-the-fly routing information cannot be obtained in advance. Unlike the recent scheme of [13] which records the entire original routing path of a data packet piggy back as the packet traverse along its path towards the sink, we propose to use WSN *routing topology tomography* [e.g., 15-19] to obtain the dynamic routing information needed for the interplay between the compressed sensing and routing. Because the overhead of routing tomography techniques is usually very small per packet compared with the recording of the raw path trace, our idea can further improve the energy efficiency of WSN compressed data collection by significantly reducing the

4overhead of path recording carried in each packet. For example, the Routing Topology Recovery (RTR) introduced in [18] only has a fixed four-byte overhead of path measurement per each packet independent of the actual path length of the packet. More importantly, the small fixed size of path measurement overhead means that it is *scalable* for large-scale WSN deployments with very long paths, as the widely used IEEE 802.15.4 communication protocol in WSNs has only the maximum size of 127-byte MAC frame including the header.

### 4.2 Measurement Matrix

Two fundamental components of CS are the random measurement (i.e., projection) matrix and the representation basis. We construct the $M \times N$ ($N = n - 1$) measurement matrix $\Phi = \{\varphi_{i,j}\}$ ($1 \leq i \leq M, 1 \leq j \leq N$) using the dynamic WSN routing matrix, leveraged by emerging WSN routing tomography technique, in our CS approach. After $M$ data packets are received in a WSN data collection cycle, the routing paths for those $M$ packets are first reconstructed via an adopted routing topology reconstruction algorithm (e.g., RTR [18]). If node $j$ is on the path of packet $i$ received at the sink, then $\varphi_{i,j} = 1$; otherwise, $\varphi_{i,j} = 0$. The $i$-th row of the measurement matrix $\Phi$ represents the routing path of packet $i$ received at the sink in the given cycle, as illustrated in equation (4).

*Proposition:* Let $G(V, E)$ be a WSN with an upward routing matrix $\Phi$ for a given data collection cycle. Suppose that $B(V, P_\Phi, H)$ is a bipartite graph with bi-adjacency matrix $\Phi$. It is feasible to use routing matrix as measurement matrix in compressed sensing in recovering $k$-sparse sensor signals in the given data collection cycle, while the expected estimation error is bounded by equation (11) in [23].

*Proof:* The problem here is an isomorphism problem of the network link delay estimation via CS in [23]. First, the sink and the leaf nodes are boundary nodes, while the others are intermediate nodes. Thus an upward routing path from a leaf node to a/the sink in the sensor network is equivalent to an end-to-end path in the network considered in [23]. Second, in our setting, the routing matrix $\Phi$ (i.e., measurement matrix) is defined in terms of the traversed nodes in each path in the network rather than the traversed links in each path defined in [23]; consequently, the formation of bipartite graph $B(V, P_\Phi, H)$ in our setting is based on the network node set $V$ as opposed to the bipartite graph $G(E, R, H)$ based on the network link set $E$ in [23]. All the theorems of [23] on the derivation of the error bounds would still be held when the $G(E, R, H)$ in [23] is replaced by our $B(V, P_\Phi, H)$. Thus the sensor signals on the nodes can be feasiblely recovered using LP as the delays on the links in [23]. We note that, due to various random noises and interferences in outdoor WSN *in situ*, the constructed measurement matrix $\Phi$ for a different data collection cycle would be quite different as a result of wireless link dynamics, even if the deployed routing protocol does not induce any additional random effect on the routing topology.

As one can see, interplaying with WSN dynamic routing on the fly, each sensor node neither stores the matching column in the measurement matrix, nor performs vector multiplication and vector addition. As sensor nodes of outdoor WSNs are usually battery-powered with very limited memory and low cost microcontroller, the compressed sensing which can effectively interplay with routing is particularly feasible and suitable to multi-hop and dynamic WSN deployments *in situ*.

### 4.3 Representation Basis

There are two main criteria in selecting a good representation basis $\Psi$: (1) its corresponding inverse has to sufficiently sparsify the signal $x$; and (2) it has to be sufficiently incoherent with the measurement matrix $\Phi$. A long-standing open question of compressed sensing for WSN data collection in conjunction with routing is how to find an appropriate representation basis $\Psi$ with good sparsification and incoherence properties [7]. To address this problem, we consider to build a suitable $\Psi$ based on graph wavelets via deep learning by [20], since the sensor data collected from a WSN are signals defined on the graph of the WSN deployment topology. Rustamov and Guibas [20] recently introduced a machine learning framework, referred to as the GDL in this paper, for constructing graph wavelets which is expected to sparsely represent a given class of signals on irregular graphs. The basic idea is to use the lifting scheme as applied to the Haar wavelets. Their insight is that the recurrent nature of the lifting scheme gives rise to a structure resembling a deep auto-encoder network. One unique advantage of their framework is the constructed wavelets are adaptive to a class of signals on the underlying irregular graph, which can better explore the inherent multi-resolution structure of a given class of signals on the underlying graph.

For any signal $f$ on graph $G$ and any level $l_0 < l_{max}$, the wavelet decomposition can be expressed as

$$f = \sum_i a_{l_0, i} \phi_{l_0, i} + \sum_{l=l_0}^{l_{max}-1} \sum_i d_{l,i} \psi_{l,i}.$$

The coefficients $a_{l,i}$ and $d_{l,i}$ are called approximation and detail (i.e., wavelet) coefficients, respectively. In GDL [20], the construction of wavelets is based on the lifting scheme, as illustrated in Figure 2.

Starting with an Haar transform (HT) defined in [20], and $l = l_{max} - 1$, $a_{l_{max}} = f$, the lifted wavelets can be obtained by iterating the process in Figure 2, where $\tilde{a}_l$ and $\tilde{d}_l$ denote the vectors of all approximation and detail coefficients of the original HT transform respectively at level $l$.

Given $n$ training functions $\{f^n\}$, the linear operators $U_l$ and $P_l$ can be learned by solving the minimization problem [20]

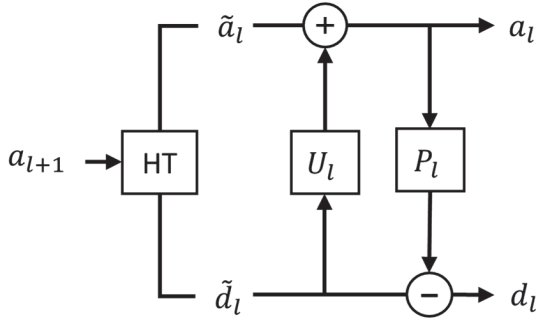

Fig. 2. Lifting scheme with Haar transform (HT). $a_l$ and $d_l$ denote the vectors of all approximation and detail coefficients of the lifted Haar wavelet transform, respectively, at level $l$. $U_l$ and $P_l$ are linear operators.

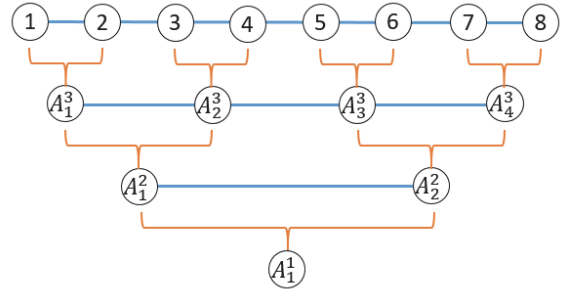

Fig. 3. An illustration of multiresolution decomposition of 1-dimensional space into a hierarchical structure, where $A_i^l$ denotes a connected segment at level $l = 1, \ldots, l_{max}$ of the illustrated linear graph (including nodes 1, 2, …, 8).

$$\{U_l, P_l\} = arg \min_{U_l, P_l} \sum_n z\left(\tilde{d}_l^n - P_l(\tilde{a}_l^n + U_l \tilde{d}_l^n)\right), \quad (7)$$

where $z$ can be any sparse penalty function. Then the representation basis can be obtained by running the inverse process of Figure 2.

However, the GDL framework [20], as expressed in (7), does not address the problem of how to generally decompose an irregular underlying graph for constructing wavelets, but rather assumes that such a hierarchical decomposition of the underlying irregular and connected graph into connected regions is already provided in advance for use in (7). Indeed, it is nontrivial to find an appropriate hierarchical decomposition of any highly *irregular* large-size graph in a general way. To overcome this difficulty, we devise a novel and effective algorithm that enables the partitioning of the multiscale structure imposed on the underlying irregular graph.

Our idea is to first embed the underlying irregular and connected graph into a linear graph (i.e., 1-dimensional space), in which any two consecutive vertices in this 1-dimensional space are, desirably, connected in the original graph. Then, signals on the original underlying irregular graph are now defined on the 1-dimensional regular space. Therefore, a standard multiresolution decomposition, such as the tree algorithm introduced by Mallat [21], can be readily applied (as illustrated in Figure 3) to generate a feasible hierarchical structure of signals on this transformed linear embedding graph, an approximate of the original underlying graph, upon which HT and (7) can be applied.

### 4.3.1 GLE Algorithm

Our devised algorithm, referred to as Graph Linear Embedding (GLE), is presented as follows. We consider the problem as finding a walk path which goes through all the vertices on the irregular and connected graph in an optimal way to reserve vertices' neighborhood information. This problem can be formulated as getting a labeling of vertices which would closely reflect the structure of the graph. This question can be related to a graph labeling problem known as the cyclic bandwidth sum problem. It consists in finding a labeling of the vertices of an undirected and unweighted graph with distinct integers such that the sum of (cyclic) difference of labels of adjacent vertices is minimized [22]. Given an undirected and connected graph $G(V, E)$ with vertex set $V = \{u_i \in V | 1 \le i \le n\}\}$ and edge set $E$, our GLE algorithm is given as follows.

1. Sort the vertices $\{u_1, u_2, u_3, \cdots, u_n\}$ in the ascending order of their degrees, so that $d_{u_1} \le d_{u_2} \le d_{u_3} \le \cdots \le d_{u_n}$. Initialize list $A = \{u_1\}$, list $B = \{u_2, u_3, \cdots, u_n\}$, and stack $C = \{u_1\}$. List $A$ keeps the vertices which have already been visited along the walk, while list $B$ keeps the remaining ones not traversed yet. The *current* vertex is defined as the most recently added vertex into stack $C$. Stack $C$ maintains the current walk path segment from the bottom vertex in the stack to the current vertex on top of the stack for further check.

2. Search a vertex $u_j$ in list $B$ that matches the following conditions: (i) $u_j$ is adjacent to the current vertex $u_i$; and (ii) $u_j$ has a neighborhood that is the most similar to the one of $u_i$. Let $Adj(u)$ return the all adjacent vertices of the vertex $u$. The similarity index between vertices $u$ and $v$, denoted as $J(u, v)$, is defined by [22]:
$J(u, v) = \frac{\#(Adj(u) \cap Adj(v) \cup \{u,v\})}{\#(Adj(u) \cup Adj(v))}$.
In another word, we are searching for vertex $u_j$ in $B$, which satisfies:
$u_j = arg \max_v J(u_i, v)$, s.t. $u_j \in Adj(u_i)$.

3. If such a vertex $u_j$ in $B$ is found, add $u_j$ into list $A$ and then delete it from list $B$. Push $u_j$ to stack $C$. If no vertex in $B$ is found adjacent to the current vertex $u_i$, pop $u_i$ out from stack $C$. Assume $u_k$ is the top element in stack $C$ now, the current vertex. Add $u_k$ into list $A$ again.

4. Repeat steps (2)-(3) until $B$ becomes empty. The ordered sequence of vertices in $A$ then forms the embedded 1-dimensional linear topology structure of the given irregular graph.

When a walk is generated by the GLE algorithm for the given connected graph, any two consecutive vertices in the resulting 1-dimensional topology structure are connected in the original graph.

*4.3.2 Construction of Underlying Graph*

Given a WSN deployment, an important consideration is how to construct the underlying graph of sensor signals from which an appropriate representation matrix $\Psi$ can be obtained in our CS approach interplaying with WSN routing. We start with the routing topology recovered at the sink for each data collection cycle in the WSN, which forms a routing topology graph (RTG). Change each directed edge in the RTG to an undirected edge, we have the corresponding undirected RTG, denoted as URTG. To maximize the incoherency between $\Psi$ and $\Phi$, we consider to construct the underlying graph as the complement graph (CG) of the WSN URTG in building our sparse representation basis $\Psi$ based on graph wavelets via deep learning.

Let P training datasets be collected from the WSN deployment for constructing $\Psi$. A training dataset corresponds to a URTG graph $G_i = (V, E_i), i \in \{1,2,...,P\}$. The union of these P graphs is

$$G_U = (V, E_U), \tag{8}$$

where $E_U = E_1 \cup E_2 \cup ... \cup E_P$. The complement graph CG of $G_U$ is

$$G_{CG} = \overline{G_U} = (V, E_{CG}), \tag{9}$$

where $(i,j) \in E_{CG}$, if and only if $(i,j) \notin E_U$.

The $G_{CG}$ is the constructed underlying graph from P WSN URTGs from training datasets for building our sparse representation basis $\Psi$, whose Laplacian matrix $L_{CG}$ will be needed to build the wavelets [20]. For an undirected graph $G = (V, E)$ along with a weight function $w: E \to \mathbb{R}^+$, where $\mathbb{R}^+$ denotes the set of positive real numbers, the adjacency matrix $A_G$ of $G$ is:

$$A_G(i,j) = \begin{cases} w(i,j) & \text{if } (i,j) \in E, \\ 0 & \text{otherwise.} \end{cases}$$

The degree matrix $D_G$ of a weighted graph $G$ is a diagonal matrix such that

$$D_G(i,i) = \sum_j A_G(i,j).$$

The Laplacian matrix $L_G$ of a weighted graph $G$ is defined as

$$L_G = D_G - A_G.$$

Let the weight of all the edges be equal to 1, then the adjacency matrix $A_{CG}$ of the complement graph $G_{CG}$ is

$$A_{CG}(i,j) = \begin{cases} 1 & \text{if } (i,j) \in E_{CG}, \\ 0 & \text{otherwise.} \end{cases}$$

And $D_{CG}(i,i) = \sum_j A_{CG}(i,j)$. We have the Laplacian matrix of the complement graph of routing as

$$L_{CG} = D_{CG} - A_{CG},$$

which will be used to find the sparse representation basis.

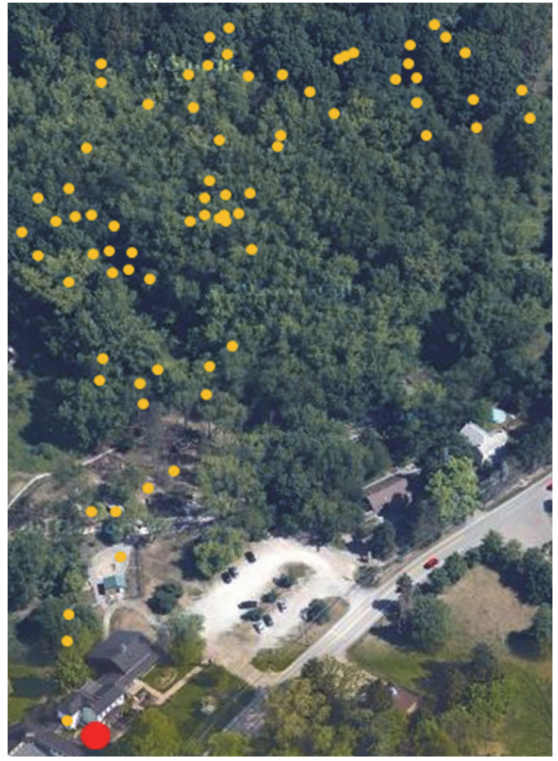

Fig. 4. WSN testbed, sink locates at the red spot in the cabin house.

## 5 VALIDATION IN REAL-WORLD WSN DEPLOYMENT

To rigorously validate the proposed CS approach based on WSN routing topology tomography, we deployed our approach in a real-world outdoor multi-hop WSN for environmental monitoring [3]. Section 5.1 describes the WSN deployment and our experiment setup. In Section 5.2, we study the sparsification performance of the constructed representation basis. In Section 5.3, we evaluate the data recovery performance of our CS approach, in comparison with three existing CS schemes CDG [8], RS-CS [7] and CDC [13].

### 5.1 WSN Testbed and Dataset

The multi-hop WSN *in-situ* used in our validation has been deployed in Pennsylvania [3], for the need of ground-based data measurements for calibrating and validating scientific models in hydrology research [24]. When our CS approach was deployed and validated on this WSN testbed, near 80 heterogeneous WSN nodes (including MICAz, IRIS, and TelosB) had been equipped and deployed forming a multi-hop network. The only energy source available for each node is provided by three NiMH AA rechargeable batteries with a nominal capacity of 2700mAh. The WSN deployment is shown in Figure 4.

TABLE 1
THE STATISTICS OF PER-PACKET ROUTING PATH RECOVERY
BY RTR SCHEME IN THE WSN TESTBED

| Total cycles | 87 |
|---|---|
| Average path recovery ratio | 98.38% |
| Best cycle recovery | 100% |
| Worst cycle recovery | 93.20% |

In terms of hardware of the WSN testbed, MicaZ [25] and IRIS motes [26], each one equipped with an MDA300 data acquisition board [27]. The MDA300 provides embedded temperature and humidity sensors. MicaZ and IRIS have 4K bytes and 8K bytes memory, respectively. TelosB motes with embedded temperature and humidity sensors. Both TelosB motes and MDA300 acquisition board have ADCs for external sensors, e.g. EC5 (soil moisture sensor) [28].
The base station (sink node), is an IRIS mote with a permanent power supply connected to a computer operated as the WSN gateway [29] where the WSN gateway forwards the sensed data stream to the WSN data management system over the Internet [30].

Our validation experiments were developed in TinyOS 2.1.2 [33]. The deployed routing protocol at the WSN testbed was an extended Collection Tree Protocol (CTP) [34], which introduces a random component into the process of packet forwarding to achieve a better traffic and energy balance. In our CS validation, we have adopted and deployed the Routing Topology Recovery (RTR) scheme proposed by Liu et al. [18]. In each data collection cycle from $M$ packets received at the sink, the RTR scheme reconstructs each per-packet routing path. In the RTR implementation, four bytes were used for carrying path measurement independent of the actual hop counts of the routing path, piggy back to each data packet routed towards the sink.

Data collected in one acquisition cycle of the WSN testbed, where each WSN node sampled and sent its sensor readings once, form a dataset. In our CS validation, we collected datasets from the outdoor WSN testbed *in situ* for 87 cycles. Table 1 shows the statistics of per-packet routing path recovery by RTR in our validation experiments conducted on the WSN *in situ*. The longest routing path of packet had 8 hops.

The first 10 collected datasets were used as the training datasets for constructing the representation basis in our approach while the remaining 77 datasets were used as the test datasets for data recovery performance. Humidity data were collected from 75 nodes, while soil moisture data were collected from 48 nodes equipped with external soil moisture sensor EC5. The original sensor readings of each node were also collected in the same data packet in each cycle in addition to the compressed sensor data to provide the base for the accuracy analysis of our CS approach.

### 5.2 Performance of Representation Basis

We first evaluate the sparseness of the representation basis. As described in Section 3, $s$ is the $N \times 1$ coefficient vector in the $\Psi$-domain with $\|s\|_o = k$, where $k \ll N$. By keeping only the largest $k$ components in magnitude in $s$, we can get the approximation $s'$ of $s$, and thus the approximation $x' = \Psi s'$. Comparing $x'$ with $x$ gives the performance of the representation basis $\Psi$.

The representation basis in our CS approach is constructed as follows: First, construct the underlying graph of WSN based on Equations (8) and (9) with recovered WSN URTGs from path measurements in given training datasets; second, use our devised GLE algorithm to obtain an appropriate hierarchy decomposition of the underlying graph; third, apply GDL [20] to the obtained hierarchy decomposition of the underlying graph to construct graph wavelets with given WSN training datasets, and then construct the sparse representation basis based on the constructed graph wavelets. Figure 5 shows an example of the humidity data collected by 75 nodes and the corresponding transform coefficients for the representation basis obtained by our approach with the 10 humidity training datasets. Similarly, Figure 6 shows an example of the soil moisture data from 48 nodes and the corresponding transform coefficients for the representation basis constructed by our approach with the 10 training datasets of soil moisture. As we can see, only very few coefficients are significant in the transform domain.

Next, we select the largest $k$ transform coefficients in magnitude of both humidity and soil moisture data, respectively, to evaluate the sparsification performance of our representation basis. We further compare the sparsification performance of our CS representation basis with those adopting Haar wavelet transformation and DCT (Discrete Cosine Transformation), the two popular transformations used in existing CS schemes such as CDC [13] and CDG [8]. The approximation error (%), defined as in (10), is employed to evaluate the performance for different CS approaches.

$$\text{Error} = \frac{\sqrt{\sum_N (x'-x)^2/N}}{\sqrt{\sum_N x^2/N}} \times 100\%. \qquad (10)$$

Figures 7 and 8 show the average sparsification errors of the 77 test datasets for humidity and soil moisture signals, respectively. As we can see, the representation basis constructed by our method (GLE+graph wavelets via deep learning by GDL) can always lead to very small approximation error even when only keeping a few largest transform coefficients in magnitude. The performances of the three different bases are improved when $k$ becomes larger, and our representation basis al-ways significantly outperforms the Haar and DCT transformations. Our method also has very stable performance on both humidity and soil moisture datasets. For humidity data, when keeping only the largest 3 transform coefficients in magnitude (out of total 75), the approximation error is less than 3.3%, while for soil moisture data, keeping only the largest 2 transform coefficients in magnitude (out of total 48), the approximation error is always less 1.7%. This indicates that the humidity and soil moisture signals can be well sparsely represented using their respective basis obtained by the proposed method.

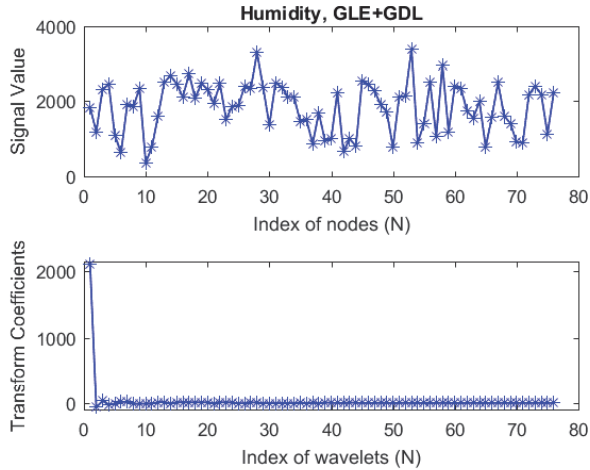

Fig. 5. Humidity raw signals (top) and the corresponding transform coefficients for our constructed representation basis using the GLE+GDL algorithm (bottom).

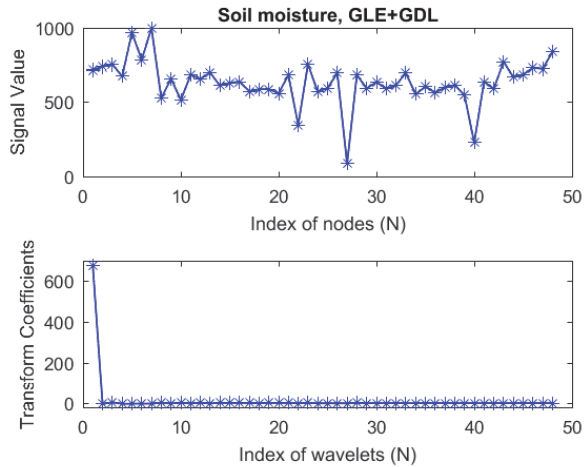

Fig. 6. Soil moisture raw signals (top) and the corresponding transform coefficients for our constructed representation basis using the GLE+GDL algorithm (bottom).

### 5.3 Signal Recovery Accuracy

After collecting $M$ ($M < N$) measurements $y =$

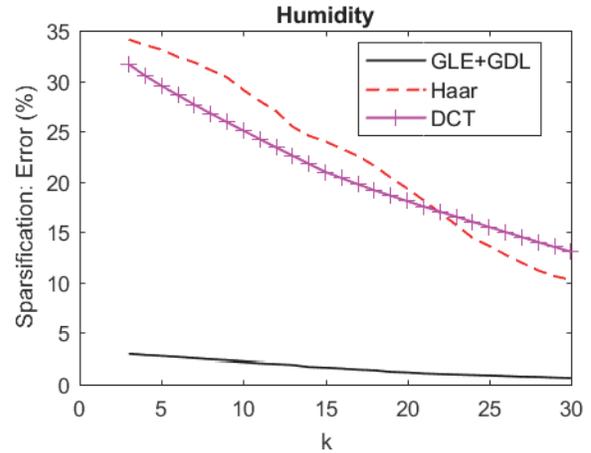

Fig. 7. The sparsification error of humidity datasets estimated by using only the largest k components in magnitude in s.

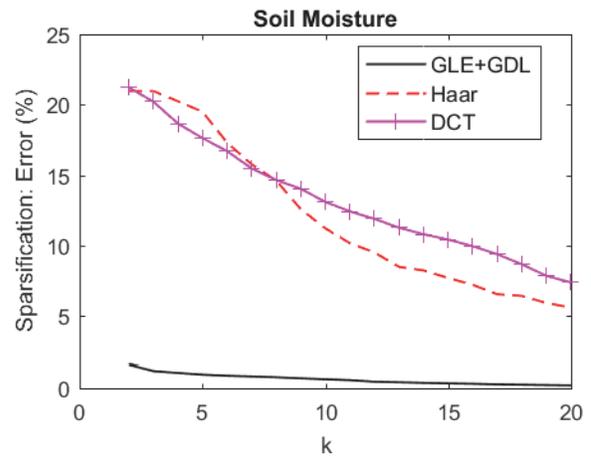

Fig. 8. The sparsification error of soil moisture datasets estimated by using only the largest k components in magnitude in s.

$[y^1, y^2, ..., y^M]^T$ from the WSN testbed in each cycle, we first recover the routing path of each received packet using the RTR scheme [18], from which the measurement matrix $\Phi$ is reconstructed for the sensor dataset in this cycle. Then, two CS solvers SL0 [35] and LP [36] are adopted to obtain an approximation $s'$ of $s$ in the transform domain. Finally, the original signal is recovered by computing $x' = \Psi s'$. The recovery accuracy is evaluated here using the approximation error defined in (10).

For the evaluation of our compressed sensing approach CSR, three existing CS schemes CDG [8], RS-CS (with *Horz-diff* transformation) [7] and CDC [13] are used for the comparison. While CSR, CDC and RS-CS approaches interplay with routing, CDG does not, and relies on dense random projections which need to collect data from all WSN nodes.

We first set $M = 12$ for both humidity and soil moisture data collection in our CS approach CSR. Figure 9

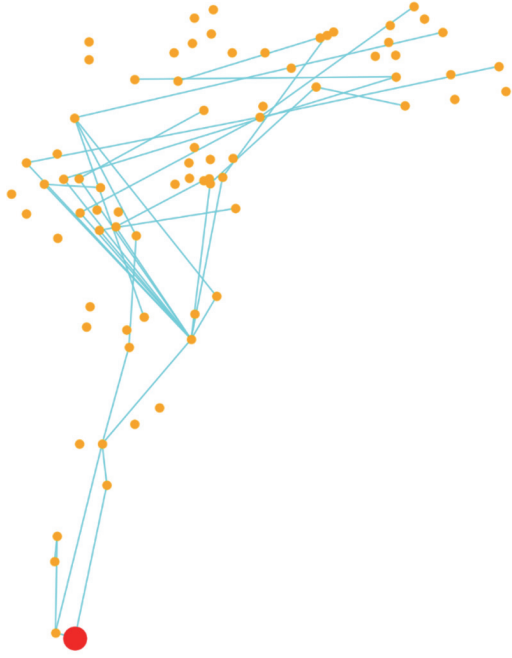

Fig. 9. An example of the routing path (measurement) topology in a data collection cycle for M=12.

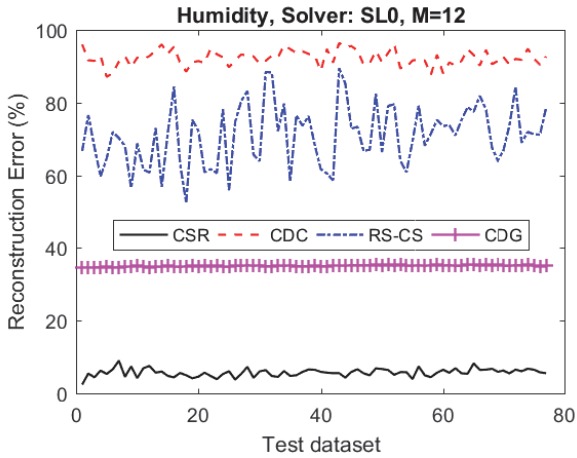

Fig. 10. Humidity dataset reconstruction error when M=12, using SL0 as the solver.

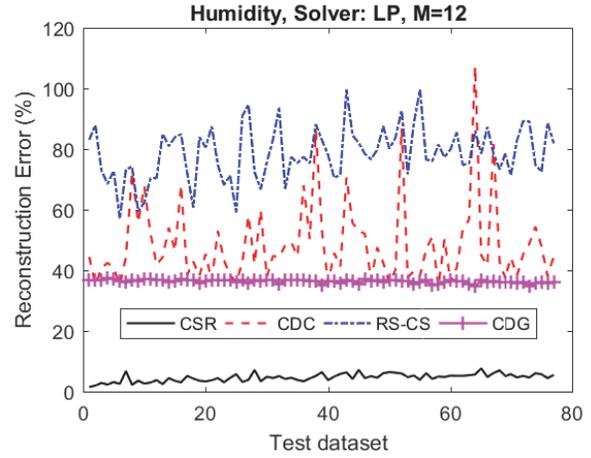

Fig. 11. Humidity dataset reconstruction error when M=12, using LP as the solver.

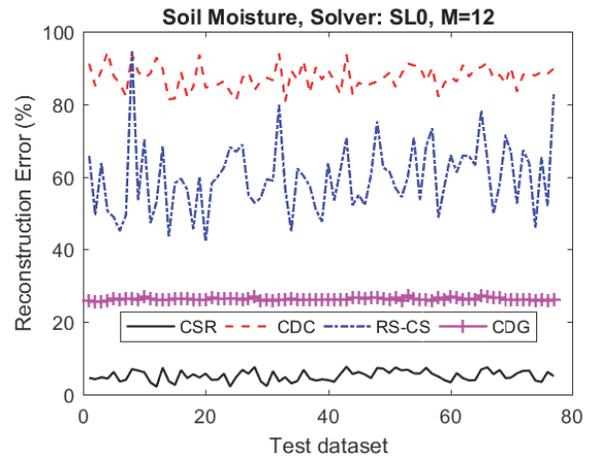

Fig. 12. Soil moisture dataset reconstruction errors when M=12, using SL0 as the solver s.

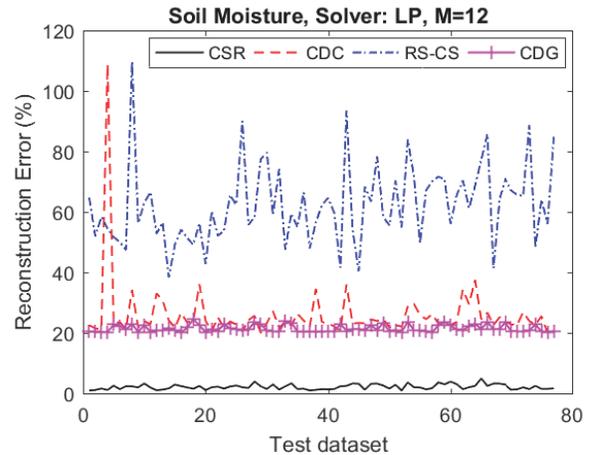

Fig. 13. Soil moisture dataset reconstruction errors when M=12, using LP as the solver.

shows an example of the WSN routing topology for the 12 measurements collected from the WSN testbed in our experiments, which was reconstructed by the RTR scheme [18] running at the sink. As we can see, many nodes were not visited in this cycle, which means their data were not collected in CSR, CDC and RS-CS approaches jointly with routing. Fewer visited nodes generally can make it more difficult to recover the entire data field.

Figures 10 ~ 13 show the reconstruction error for humidity and soil moisture signals using four different CS approaches, with two solvers: SL0 and LP. As we can see,

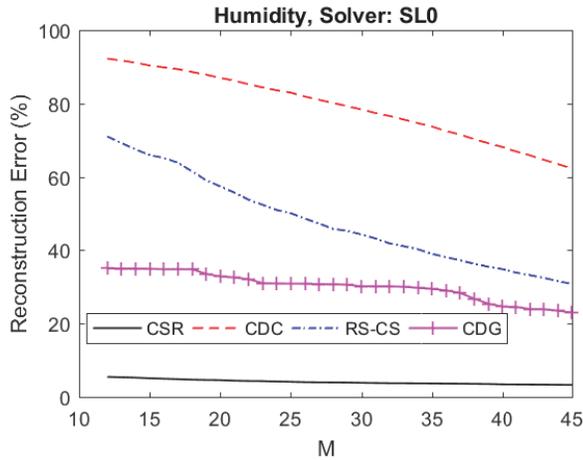

Fig. 14. Humidity dataset reconstruction errors with different M, using SL0 as the solver.

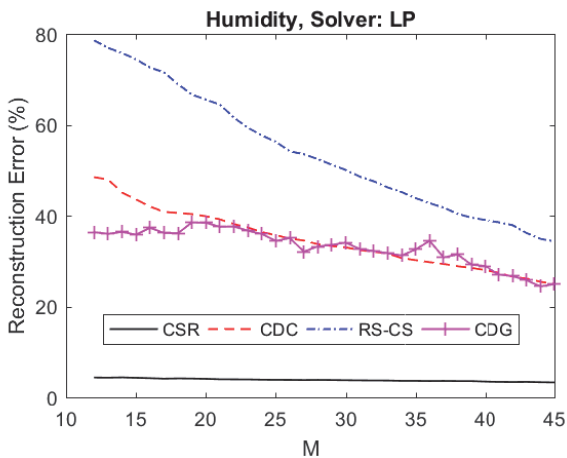

Fig. 15. Humidity dataset reconstruction errors with different M, using LP as the solver.

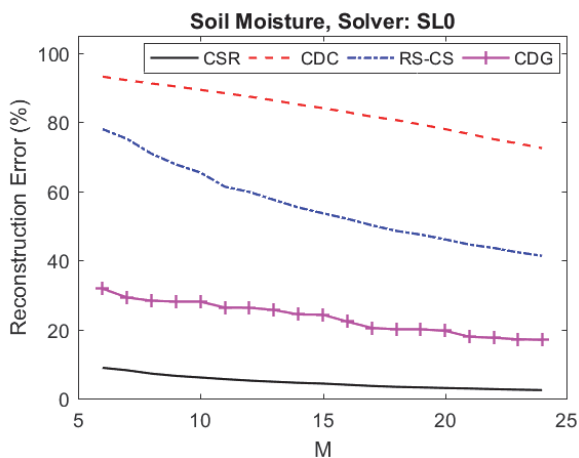

Fig. 16. Soil moisture dataset reconstruction errors with different M, using SL0 as the solver.

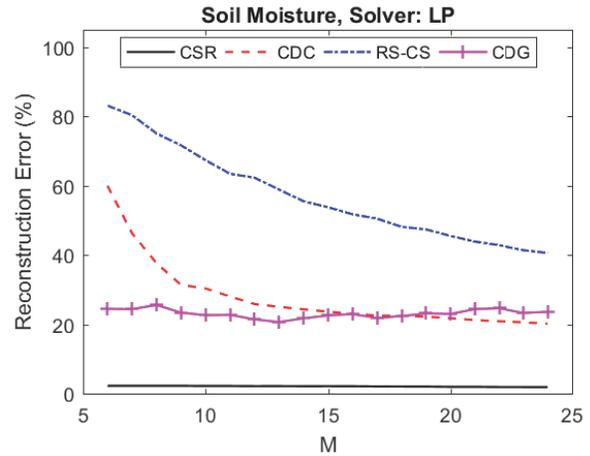

Fig. 17. Soil moisture dataset reconstruction error with different M, using SL0 as the solver.

our CSR with LP solver can always achieve the data recovery with the error less than 7.7% on humidity datasets and less than 5.0% on soil moisture datasets for any collection cycle, significantly outperforms CDG, CDC and RS-CS. Note that both CDC and RS-CS perform worse than CDG and they are sensitive to different datasets collected in different cycles, because only a subset of the nodes is visited in each individual collection cycle. Our CSR approach overcomes this problem by constructing a much better representation basis as shown in Section 5.2, therefore CSR always outperforms the other three CS approaches for both humidity and soil moisture data. We also note that CDC are very sensitive to the solver, it performs much better with LP than SL0. Generally, the solver LP outperforms SL0, but LP takes longer computation time.

Figures 14 ~ 17 show the reconstruction errors for humidity and soil moisture signals using four different CS approaches, with different numbers ($M$) of collected measurements. As we can see, our CSR has excellent performance even when $M$ is very small, with reconstruction errors being an order of magnitude less than those of the other three CS approaches even with much larger $M$. Generally, the performance will improve when $M$ becomes larger, the only exception is CDG with solver LP on soil moisture data.

Figure 18 gives an example of the reconstructed humidity data field by our CSR when only 12 data packets are collected at the sink, in comparison with the original humidity data field. As we can see, while the original humidity data change drastically from sensor node to node, our CSR is still able to recover the entire data field with high fidelity using only 12 measurements.

Figures 19 and 20 show the transmission numbers of CSR and CDG for collecting humidity and soil moisture measurements, respectively, for different numbers of received measurements. As we can see, the CSR leads to the drastic reduction of data packet transmission numbers, by an order of magnitude less than those of CDG. This means the drastic radio communication energy conservation by CSR, the great advantage of any CS approach successfully interplaying with routing. While both CDC and RS-CS have the same transmission numbers as CSR in our experiments, due to the employment of the same routing protocol CTP+EER in the experiments for CDC and RS-CS as well, CDC and RS-CS have significantly larger data packet size than that of CSR. This is because CDC and RS-CS record the original packet path in the data packet along the route, whereas our CSR uses routing topology tomography for path information. Consequently, CSR is not only scalable for large-size WSNs and big data acquisition, but also more resource efficient than CDC and RS-CS.

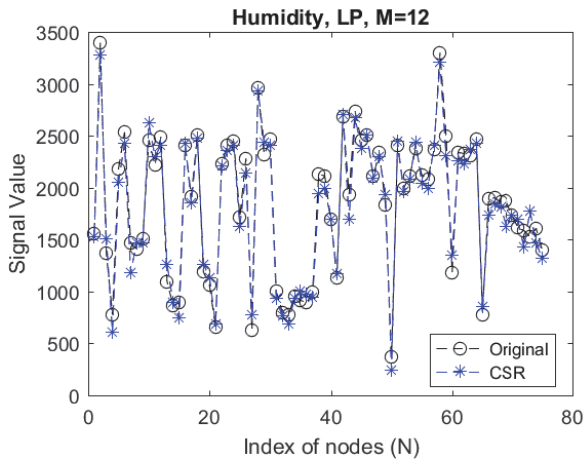

Fig. 18. Original signals vs. the reconstructed signals by CSR.

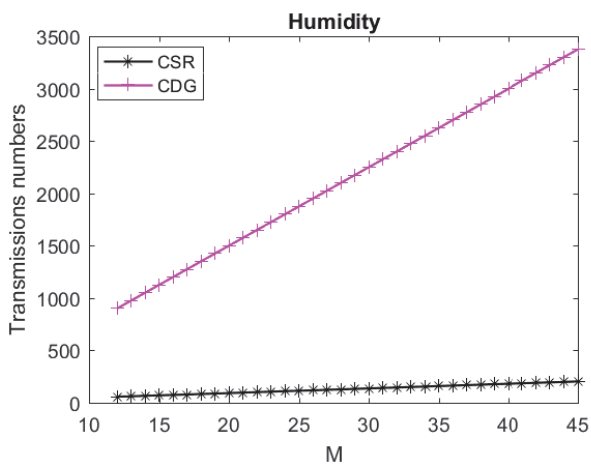

Fig. 19. Transmission numbers on different approaches on humidity dataset.

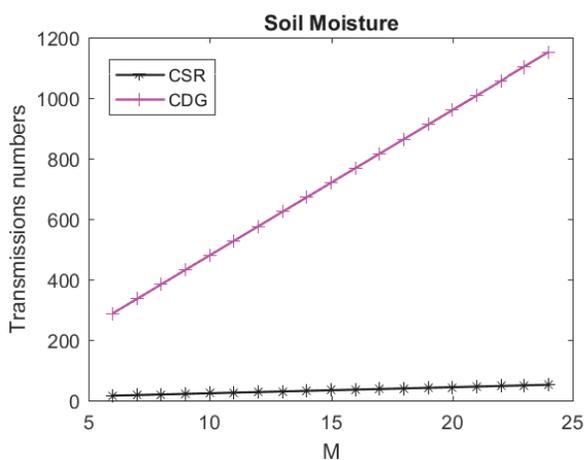

Fig. 20. Transmission numbers on different approaches on soil moisture dataset.

In summary, rigorous validation and evaluation of our CSR approach versus three existing CS approaches CDG, CDC and RS-CS were conducted in a real-world WSN deployment *in situ*. The results clearly demonstrate that CSR approach significantly outperforms CDG, CDC and RS-CS by reducing data reconstruction errors by an order of magnitude for the entire WSN data field, while drastically reducing wireless communication costs, by an order of magnitude, at the same time. This indicates that our CSR approach is a reliable and practical solution to energy efficient data acquisition in multi-hop large-scale WSNs. In our experiments, CSR can successfully recover the entire data field of the real-world multi-hop WSN *in situ* with very small errors, when only 16% of data packets (i.e., 12 randomly selected nodes out of total 75 sensor nodes in the WSN testbed) needed to be collected at the sink.

## 6 CONCLUSION

In principle, CS based data acquisition in multi-hop WSN deployments has a great potential to further significantly reduce sensor nodes' transmissions via the interplaying with the network dynamic routing to facilitate wireless big data acquisition. In practice, however, two critical issues have to be effectively addressed before the potential said above can be realized in any large-scale real-world WSN deployment. The first issue is how to effectively obtain the dynamic routing information for each received packet at the sink, since simply recording path along the route is neither scalable nor resource-efficient. The second issue, originally identified by Quer et al. [7], is how to design a suitable representation basis for real-world signals that has good sparsification and incoherence with the measurement matrix derived from dynamic WSN data packet routing, because it has been found [7] that commonly used transformations including DCT and Haar Wavelet for constructing representation basis, as well as the Horz-diff transformation [7], all suffered from large recovery errors for real WSN data.

In this paper, we attempt to address these critical open questions and present a novel CS approach called CSR for multi-hop WSN data acquisition based on dynamic routing topology tomography. Our CSR approach has two distinguishing characteristics. First, CSR introduces the use of WSN routing topology tomography into CS approach and thus provides a practical and elegant solution for large-scale WSN data acquisition based on effective interplaying with dynamic routing. We show that the adoption of routing matrix as measurement matrix in compressed sensing in recovering *k*-sparse sensor signals in WSN can achieve feasible estimation with bounded errors. As shown in our real-world WSN experiments, our CSR approach considerably reduces transmission numbers (e.g., 58 transmissions in CSR versus 900 transmissions in CDG, for collecting 12 measurements at the sink), resulting in an order of magnitude less in energy consumption compared to CDG, and also significant reduction of transmission costs compared to CDC and RS-CS, thus extending the lifetime of real-world outdoor WSN deployments. Second, CSR provides a systematic method to construct an optimized representation basis with both good sparsification and incoherence properties for various given classes of signals, and therefore drastically reduces WSN data recovery errors by an order of magnitude compared to existing CS schemes CDG, CDC and RS-CS. Therefore, the proposed approach is expected to significantly improve the state of art of CS based approaches for WSN data acquisition, and to facilitate the CS application in large-scale multi-hop outdoor WSN systems for various data gathering.

Our approach is deployed for a real-world outdoor WSN testbed and is rigorously validated and evaluated via the WSN deployment *in situ* operated under highly dynamic communication environment for environmental monitoring. To the best of our knowledge, our work represents the first demonstration and performance analysis of CS approaches applied to real-world WSN deployment *in situ* jointly with routing for data acquisition with actual routing protocol in operation. Our approach via deep learning seems very effective, as only 10 training datasets were used in constructing the representation basis in our experiments.

It is expected that the presented systematic method in our CSR approach for constructing an optimized representation basis can be in general adopted to any other CS schemes to significantly improve their data recovery fidelity in big data acquisition.

## ACKNOWLEDGMENT

The authors wish to thank X. Zhong at IUPUI, and G. Villalba, F. Plaza, T. A. Slater and Dr. X. Liang at the University of Pittsburgh for their great help and support in the WSN testbed deployment of the CSR approach. This work was supported in part by the U.S. National Science Foundation under CNS-1320132.